# Dialog Management for Decision Processes

*Paul Ioan Fodor*

Computer Science Department, Stony Brook University, Stony Brook, NY 11720, USA
pfodor@cs.sunysb.edu

## Abstract

The search for a standardized optimum way to communicate using natural language dialog has involved a lot of research. However, due to the diversity of communication domains, we think that this is extremely difficult to achieve and different dialogue management techniques should be applied for different situations. Our work presents the basis of a communication mechanism that supports decision processes, is based on decision trees, and minimizes the number of steps (turn-takes) in the dialogue. The initial dialog workflow is automatically generated and the user's interaction with the system can also change the decision tree and create new dialog paths with optimized cost. The decision tree represents the chronological ordering of the actions (via the parent-child relationship) and uses an object frame to represent the information state (capturing the notion of context). This paper presents our framework, the formalism for interaction and dialogue, and an evaluation of the system compared to relevant dialog planning frameworks (i.e. finite state diagrams, frame-based, information state and planning-based dialogue systems).

## 1. Introduction

Dialogue is the fundamental communication mechanism. Most of the research involving dialogue management for artificial systems has been focused on the search for a common formalism that can implement all possible formalisms and, at the same time, to be optimal regarding a specific metric (e.g. duration of dialogue, number of turn-takes, user's satisfaction). Our work focuses on a specific set of problems that can be represented using decision trees. This extends mix-initiative dialog systems with learning a tree of probabilistic optimal dialog paths from a set of previous dialog paths executed by humans. The metric that we want to optimize in this case is the number turn-taking steps in the dialogue. In order to evaluate our approach, we implemented a framework that acts as a credit card application screening manager. We use an initial dataset collected by human operators to construct an initial decision tree. The dataset represents the positive and the negative instances of people who were or were not granted credit based on a set of different tuples of attributes with values: period lived at the current address, current salary, availability of a savings account, age, defaulted on a loan (yes/no), number of other credit cards that the person already has, ever declared bankruptcy, etc. Our method integrates expectations about how users will answer questions (in effect, distributions over users' responses to questions, which we see as a user model) with a model of the task (e.g. how credit decisions are made). Using decision trees to tackle this is a novel approach.

First, we present a summary of facts about human conversation, turns and utterances, speech acts, and dialogue techniques. Then, in the main section we introduce the components of our conversational language system and the formalism used to represent dialogs. In the evaluation section we present our dialogue mechanism's performance and compare it to other dialog techniques, including state machine, frame-based, information-based and planning-based techniques.

## 2. Conversational systems

In the past 30 years, artificial dialogue systems have been developed for domain directed dialogues (e.g. making travel arrangements [1], and the ATIS corpus for air traffic information systems [2]) using various techniques. A number of metrics have been considered to evaluate such dialogue systems: the user's satisfaction, the duration of interaction, and the numbers of turns involved in completing a task. Currently, the common perception is that a dialog is composed of atomic units called *speech acts* that are action performed by the speaker as utterances in a dialogue [3]. Furthermore, these speech acts are endowed with a more complex function once they are utilized in a conversation: they become speech acts with internal conversational functions (denoted as *dialogue acts* or *conversational moves* [4, 5]). These conversational moves are used by the dialog manager to resolve a goal in some domain world. The *dialogue manager* is the component that takes the input from the speech recognition (ASR) and the natural language understanding (NLU) components, reasons about an internal task, and passes the output to the natural language generation (NLG) and the text-to-speech synthesis (TTS) modules. This handling of both inputs and outputs creates the premises to conduct a dialogue with the user.

We identified various approaches of dialogue management in literature: finite state diagram based, frame-based, planning and information-state dialogue managers, and probabilistic based.

Finite-state dialogue systems are fixed finite state machines, in which the system has the initiative at each turn and the user answers exactly the question that the system asked. Obviously, it is difficult to create mixed initiative systems with finite state dialog managers because the number of states can be very large (i.e. separate states are required for each possible subset of questions that the user's statement could be addressing).

Frame-based (or form-based) dialogue managers ask the user questions to fill slots in a frame (form), but restricts the user to fill slots in the same frame only. Each slot is associated to a question (e.g. if the empty slot is "city", the question is "What is the city?"). More advanced frame systems may also allow slots to be filled in more than one frame (e.g. DARPA Communicator [6]) or to disambiguate which slot is filled by the user's response by defining sets of production rules (e.g. the Mercury flight reservation system [7]).

More advanced dialog management techniques use AI planning techniques for controlling the conversation in planning domains (e.g. SmartKom [8], TRINDIKIT [9]). This requires the conversation acts to be defined like any other action with preconditions and effects on the internal state.

One disadvantage of finite state, frame-based and plan-based dialogue management techniques is that the dialog manager has to manually develop the dialog (i.e. develop the state machines, the frames or the actions with their preconditions, generic state updates and post-conditions). Also, the structures of the dialog are fixed, meaning that the current dialogs do not optimize future dialogs. New probabilistic-based dialog management techniques were developed [10-12] to construct the dialog workflow automatically from an initial dataset of conversations for a specific domain (by using Markov Decision Processes, reinforcement learning, Partially Observable Markov Decision Processes or other statistical based methods). Most of these methods search for standardized way to represent all dialogs or a large range of conversational systems. However, due to the diversity of communication domains, we think that it is extremely difficult to achieve an optimum solution for all different dialogue management techniques. In our framework we address a simpler problem: that of decision processes for classification which minimizes the number of take-turns. We construct a decision tree completely automatic from an initial set of dialogs in a specific classification domain (e.g. credit application screening). An additional feature of our approach is that each dialog verified by a human supervisor adds its contribution to the decision tree, influencing future dialogs.

## 3. Methods

Our conversational system's architecture contains the following units: a speech recognition module, a natural language understanding module, the dialog and task completion manager, the natural language generation module and the text-to-speech synthesis module. Due to space constraints, we will describe only the dialog manager unit. Its goal is to update the dialogue context, to interface the dialog system with the external applications and databases, and to decide what dialog act to execute next.

The dialog manager is based on probabilistic decision trees [13, 14] and can solve decision problems. Our dialog manager uses an initial set of dialogs collected by human operators to learn and construct the initial decision tree. The dataset $S$ contains a set of values for the target concept associated with tuples (frames) of attributes (slots) $<A_i$, i =1,n>. A decision tree is a tree in which all childless nodes contain a classification for the target concept, the nodes with children contain a set of dialog acts in which the value for one slot (attribute) $A_i$ in the structure is inquired, confirmed and validated, and the edges are labeled with a value or a condition for the attribute $A_i$ in the parent node. The problem is finding the minimum height decision tree that classifies the target concept. The question that follows is which attribute is the best classifier at each step (i.e. which attribute provides the greatest information gain at each step)? To answer this question, we compute the information gain, $Gain(S, A)$ of all attributes $A$, relative to the collection of examples S, and we select the attribute A for which $Gain(S, A)$ is the maximum. Then, we create a node with the attribute A in the decision tree and we continue the process until we have no more attributes left (i.e. the decision path is complete). The information gain, $Gain(S, A)$ is defined as:

$$Gain(S, A) \equiv Entropy(S) - \sum_{v \in Values(A)} \frac{|S_v|}{|S|} Entropy(S_v)$$

where $Values(A)$ is the set of all possible values for attribute $A$, and $S_v$ is the subset of $S$ for which the attribute $A$ has value $v$ (i.e., $S_v = \{s \in S | A(s) = v\}$). The entropy of the set $S$ is defined as:

$$Entropy(S) \equiv \sum_{i=1}^{c} -p_i \log_2 p_i$$

where $c$ is different values that can be taken by the target argument, and $p_i$ is the proportion of $S$ belonging to class $i$.

Using the decision tree, our dialog manager is able to construct a dialog with the user following the edges corresponding to the values of the attributes contained in the responses from the user. As a result, the dialog consists of a minimal number of turns.

Each dialog can be recorded, verified and validated by a human operator. This operator can change the classification result of the target concept. As a consequence, the dialog manager will refine the decision tree automatically using standard algorithms from [13]. The modified tree can be subsequently used in future dialogs.

The problems that can occur in this process are missing values for the selection attributes (i.e. when the user cannot provide information to a specific question). In this case, we have two options: either move to the highest probability current node's child or get all paths starting from the current node and iteratively select the children with the greatest probability value. In the second case, the process will follow a parallel search of all possible future paths.

## 4. Results

To evaluate our approach, we implemented a framework which acts as a credit card application screening manager. We used the initial dataset (see Figure 1) to build the decision tree. Our dataset contained 26 different attributes: period lived at the current address, current salary, availability of a savings account, age, defaulted on a loan (y/n), number of other credit cards that the person already has, ever declared bankruptcy, etc. Due to space constraints, we simplified the decision tree in the Figure 2 of this paper to only 4 attributes.

| Employment | Years | Savings | Bankruptcy | … | Credit |
|---|---|---|---|---|---|
| no (self-employed) | 10 | 100,000 | yes | | yes |
| no | - | 5,000 | yes | | no |
| yes | 1 | 2,000 | no | | yes |
| … | | | | | |

Figure 1: *Initial dataset of credit approvals.*

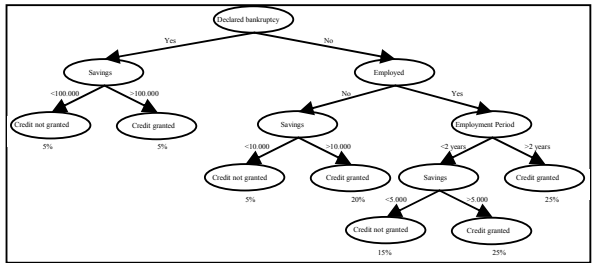

Figure 2: *Decision tree for credit screening.*

In Figure 3, we show a credit application example which lacks information. At each step the system computes the probability of all the future states and chooses the most probable node. In this specific example the system computed that the "Credit granted" classification had a greater probability than the "Credit not granted" classification.

| |
|---|
| System: Did you ever declare bankruptcy? |
| User: No. |
| System: Are you employed? |
| User: I cannot tell you that. |
| System: How much do you have in savings? |
| User: Fifteen thousand dollars. |
| … |
| System: OK. We grant your credit. |

Figure 3: *Schematic diagram of speech production.*

We implemented similar versions of the credit screening application as: a finite-state system, a frame-based system, a planning system in TRINDIKIT, and a decision tree system. Furthermore, we measured the average number of turn-takes over 20 uses in each (speech-based) system to complete a credit screening operation (see Figure 4), and the users' perception of the system (measured as the user satisfaction surveyed at the end of the interaction) (see Figure 5). Throughout all experiments we obtained results significantly better using the decision tree technique rather than the other methods (state machines, frame-based and plan-based).

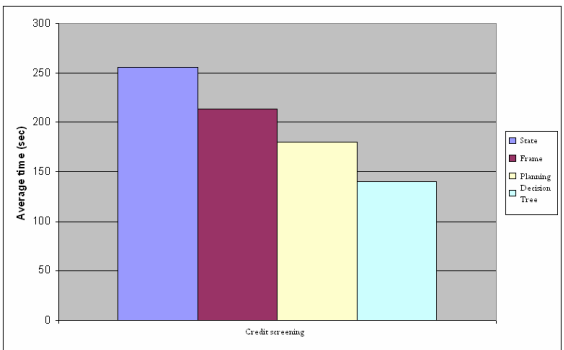

Figure 4. *Average time to complete a credit screening*

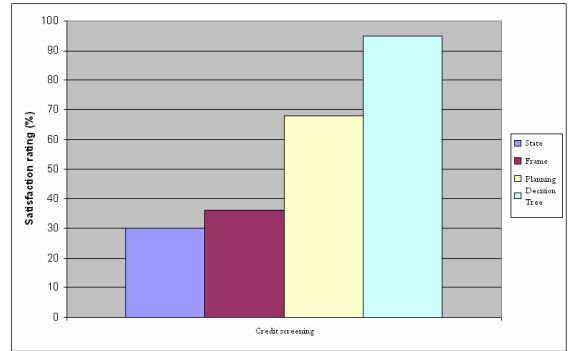

Figure 5. *User satisfaction*

## 5. Discussion

Creating the finite state machine, the frames, or the actions for planning (i.e. preconditions, generic state updates and post-conditions) manually, is a time consuming and challenging task for the dialog developer and represents a "shared" disadvantage of the finite state, the frame-based and the planning systems. Our new approach for decision processes constructs the decision tree completely automatic, based on an initial dataset of conversations.

Another disadvantage of finite state, frame-based and planning based dialog managers is that they are fixed (i.e. the current dialogs do not optimize the dialog system). As opposed to this situation, in our system, each dialogue verified and validated by a human supervisor brings its contribution to the decision tree, influencing future dialogues.

We also compared our approach with probabilistic-based dialog management techniques developed in [10,11,12]. We reached to the conclusion that we only address a simpler problem than these works (that of decision processes for classification which minimizes the number of take-turns) using a novel approach (that of decision trees), that to our knowledge was not used before for dialog management.

In all our experiments we considered the input obtained from the speech recognition engine as accurate. One problem we studied was how to include the recognition probability in our computation. A possible solution is using the decomposition of [11] of the observation into a discrete component $h$ (the speech recognition hypothesis) and a continuous component $c$ (the recognition confidence score) and using an observation function $p(h,c \mid n)$, where $n$ is a node in the decision tree, and the observation function $p$ is the probability of being in the node $n$ and recognizing a set of words $h$ with a recognition probability $c$.

One point to mention is how mixed initiative is handled in our system (i.e. what happens if the person ignores or refuses to answer a question and instead provides some other information, or how can the system make use of additional information the user provides in response to a question). We exploited a set of possibilities by constructing a partial execution tree and computing the probabilities of the dialog to be in each of these nodes. Because of the extent of this solution, we will try to document it in the future.

Another point to mention is that there is no guarantee that the decision tree will always make the right decision. Decision trees will have to *guess* when they encounter a new case (not in the training data). This guess will sometimes be wrong. A possible solution is to have an exception method, when the system could fall back to the original algorithm when it encounters a *new* application (i.e., an application with a combination of features it hasn't seen before). In this case other machine learning approaches would be more appropriate (e.g., case-based learning).

## 6. Conclusions

Our research on dialog management for decision processes shows that by learning the decision trees for conversations one can optimize the dialogue management. Furthermore, the findings of this experiment can be successfully applied in other dialog applications, such as contact center solutions or audio Web browsing.